\documentclass[aps,prl,twocolumn,groupedaddress,showpacs,floatfix]{revtex4}

\usepackage{amssymb}

\usepackage{graphicx}

\begin{document}

% Use the \preprint command to place your local institutional report
% number in the upper righthand corner of the title page in preprint mode.
% Multiple \preprint commands are allowed.
% Use the 'preprintnumbers' class option to override journal defaults
% to display numbers if necessary
%%\RCS $Id: text.tex,v 1.39 2005/01/20 15:03:24 denis Exp $
%%\preprint{\texttt{\RCSId}}

%Title of paper
\title{Dendritic to globular morphology transition in ternary alloy
  solidification}

% repeat the \author .. \affiliation  etc. as needed
% \email, \thanks, \homepage, \altaffiliation all apply to the current
% author. Explanatory text should go in the []'s, actual e-mail
% address or url should go in the {}'s for \email and \homepage.
% Please use the appropriate macro foreach each type of information

% \affiliation command applies to all authors since the last
% \affiliation command. The \affiliation command should follow the
% other information
% \affiliation can be followed by \email, \homepage, \thanks as well.
\author{Denis Danilov}
%\email[]{}
%\homepage[]{http://www.home.fh-karlsruhe.de/~dade0001/}
%\thanks{}
%\altaffiliation{}
\author{Britta Nestler}%
\affiliation{Institute of Applied Research, Karlsruhe University of
  Applied Sciences, Moltkestrasse 30, 76133 Karlsruhe, Germany}
%\homepage[]{http://www.fbi-lkt.fh-karlsruhe.de/lab/studies/pace/}

%%\date{\today}
%%\date{\texttt{\RCSId}}

\begin{abstract}
  The evolution of solidification microstructures in ternary metallic
  alloys is investigated by adaptive finite element simulations of a
  general multicomponent phase-field model. A morphological transition
  from dendritic to globular growth is found by varying the alloy
  composition at a fixed undercooling. The dependence of the growth
  velocity and of the impurity segregation in the solid phase on the
  composition is analyzed and indicates a smooth type of transition
  between the dendritic and globular growth structures.
\end{abstract}

% insert suggested PACS numbers in braces on next line
\pacs{81.10.Aj, 61.66.Dk, 68.70.+w, 81.30.Fb}
% insert suggested keywords - APS authors don't need to do this
%\keywords{}

%\maketitle must follow title, authors, abstract, \pacs, and \keywords
\maketitle

%%%%%%%%%%%%%%%%%%%%%%%%%%%%%%%%%%%%%%%%%%%%%%%%%%%%%%%%%%%%%%%%%%%%%%
%%% Introduction %%%%%%%%%%%%%%%%%%%%%%%%%%%%%%%%%%%%%%%%%%%%%%%%%%%%%
Multicomponent alloys form the most important class of metallic
materials for technical and industrial processes. Combined with the
number of components is a wealth of different phases, solidification
processes and pattern formations. The mechanical properties of a
material strongly depend on the morphology and on the characteristical
quantities of the microstructure.  One of the most common growth
morphologies in metallic alloys are dendrites. During growth,
dendritic patterns generate solute microsegregation forming the
structure of grain boundaries on a larger scale.  The growth of
dendrites from an initially undercooled metallic melt is caused by a
negative concentration gradient ahead of the solidification front.
Extensive studies have been made to explore dendritic growth in pure
substances and in binary alloys.  The constitutional undercooling and
the corresponding concentration gradient at the growing front are the
main factors leading to morphological instabilities of the
solid-liquid interface \cite{Mullins1964,Trivedi1986}.  In a binary
alloy, the relation between the temperature interval $\Delta T_0$ of
the liquidus and solidus line, the concentration of the impurity
component $c_0$ in the melt, the partition coefficient $k_e$ and the
liquidus slope $m_e$ is given by $ \Delta T_0 = - m_e c_0
(1-k_e)/k_e$, according to the equilibrium phase diagram of a
stationary planar front.  The temperature interval determines the
magnitude of the constitutional undercooling at a growing solid-liquid
interface and is hence a significant material quantity for the
stability properties.  The more pronounced concentration boundary
layer of the impurity ahead of the moving interface at larger values
$\Delta T_0$ drives the occurrence of instabilities leading to the
formation of cellular and/or dendritic structures.  For multicomponent
alloys as e.g. in ternary systems, the interface instabilities are
driven by the characteristic diffusion properties of each solute field
in the multicomponent system. Using an approximated model with
independently acting solutes, it was analytically shown in
\cite{Coriell87}, that each solute enhances morphological instability
of a solidification front with different weighting factors.  A
variation of the alloy composition causes a change of the strength of
instability forces on the growing interface, because the value of the
liquidus--solidus temperature interval strongly depends on the alloy
composition.  Investigations of the interacting quantities and of the
fundamental mechanisms during solidification of ternary or higher
component materials have just recently become more intensive by means
of numerical modeling.

To numerically treat the complex solid-liquid interface geometry of
dendritic crystals, a phase-field approach has the advantage of
avoiding the explicit tracking of the phase boundary.  The solid and
liquid are distinguished by a phase-field variable $\phi(x,t)$ with a
smooth transition from $0$ to $1$ leading to a diffuse interface. In
the sharp interface limit, classical free boundary problems and the
Gibbs-Thomson law can be recovered from the phase-field approach (see
e.g. \cite{Caginalp93} and references therein).  In the two past
decades, there has been a great progress in phase-field modeling of
growth structures in pure substances and alloy systems.  For a
historical background and applications see the review article in
\cite{Boettinger2002} and references therein.

In this Letter, the influence of changing the alloy composition on the
interface stability, on the characteristic morphology and on the
growth velocity is investigated by performing numerical simulations of
a general multicomponent phase-field model with interacting and
coupled diffusion fields.  We chose the ternary
Ni$_{60}$Cu$_{40-x}$Cr$_{x}$ alloy system as a prototype for this
study and herewith build upon intensive studies of the binary Ni-Cu
system (e.g. \cite{Warren1995} and \cite{Hoyt1999}). Hence, the
corresponding physical parameters are established relatively well.
Furthermore, the liquidus--solidus interval for binary
Ni$_{60}$Cu$_{40}$ is about three times larger than the interval for
binary Ni$_{60}$Cr$_{40}$ and one can expect a significant dependence
of growth morphology on the concentration of the alloy components.

%%%%%%%%%%%%%%%%%%%%%%%%%%%%%%%%%%%%%%%%%%%%%%%%%%%%%%%%%%%%%%%%%%%%%%
%%% Model %%%%%%%%%%%%%%%%%%%%%%%%%%%%%%%%%%%%%%%%%%%%%%%%%%%%%%%%%%%%
For the subsequent computations of ternary alloy solidification, we
apply a recently developed phase-field model to the case of 3
components and 2 phases. The general formulation of the phase-field
model in \cite{GarckeSIAM} allows for an arbitrary number of phases
and components and is based on an entropy density functional of the
form
\begin{equation}
  \label{eq:entropy_functional}
  S
  = \int_{\Omega} \left[
    s(e, c, \phi)
    - \left(
      \varepsilon a(\phi, \nabla \phi)
      + \frac{1}{\varepsilon} w(\phi)
    \right)
    \right] dx.
\end{equation}
The bulk entropy density $s$ depends on the internal energy $e$, on a
concentration vector $c=(c_1,\ldots,c_N)$ and on a phase-field vector
$\phi=(\phi_1,\ldots,\phi_M)$. The vector components $c_i$,
$i=1,\ldots,N$, represent the solute concentrations in an
$N$-component alloy.  The variable $\phi_{\alpha}$,
$\alpha=1,\ldots,M$, denotes the local volume fraction of phase
$\alpha$ in an $M$-phase system.  The thermodynamics of the interfaces
is determined by the gradient energy density $a(\phi, \nabla\phi)$, by
the multiwell potential $w(\phi)$ and by the small scale parameter
$\varepsilon$ related to the thickness of interface.  The gradient
energy and the multiwell potential depend on the surface entropy
density $\gamma$.  The evolution equations are obtained from
Eq.~(\ref{eq:entropy_functional}) in a thermodynamically consistent
way ensuring a combination of positive local entropy production and
mass conservation laws.  By variational derivatives $\partial_{t}
\phi_\alpha \sim \frac{\delta S}{\delta \phi_\alpha} - \lambda$, the
phase-field equations read
\begin{equation}
  \label{eq:phase-field}
  \omega\varepsilon \frac{\partial \phi_\alpha}{\partial t}
  = \varepsilon
  \left( \nabla \frac{\partial a}{\partial \nabla \phi_\alpha} -
    \frac{\partial a}{\partial \phi_\alpha} \right)
  - \frac{1}{\varepsilon}
 \frac{\partial w}{\partial \phi_\alpha}
  - \frac{1}{T} \frac{\partial f}{\partial \phi_\alpha}
  - \lambda,
\end{equation}
where $f(c,T,\phi)$ is the free energy density, $\lambda$ is a
Lagrange multiplier taking the constraint $\phi_1+\ldots+\phi_M =1$
for the phase fields into account and $\omega$ is a kinetic factor
related to the kinetic coefficient $\mu$.
In the sharp interface limit, it has been discussed in
\cite{GarckeSIAM} that Eq.~(\ref{eq:phase-field}) relates to the
Gibbs-Thomson equation for a moving interface.

In our numerical investigations, the alloy composition will be changed
by keeping the Ni concentration at a constant 60\,at.\% and by
exchanging Cu by Cr, i.e.  Ni$_{60}$Cu$_{40-x}$Cr$_{x}$ with
$0<x<40$\,at.\%.  Three assumptions are applied: (i)~We examine
primary dendritic growth involving only two phases in the ternary
system: a solid and a liquid.  The eutectic region in the Ni-Cr phase
diagram is not considered.  For a solid-liquid system, the phase-field
model reduces to one variable $\phi$ denoting the fraction of solid
phase.  (ii)~The system is considered in ideal solution approximation
with a free energy density of the form
\begin{equation}
  \label{eq:free-energy-density}
  f(c,\phi)
  = \sum_{i=1}^3 c_i L_i \frac{T-T_i}{T_i} h(\phi)
  + \frac{RT}{v_m}\sum_{i=1}^3 c_i \ln c_i,
\end{equation}
where $L_i$ and $T_i$ are the latent heat and the melting temperature
of component $i$, respectively, $R$ is the gas constant, $v_m$ is the
molar volume and $h(\phi)$ is a monotone function on $[0,1]$ that
satisfies $h(0)=0$ and $h(1)=1$.  (iii)~Following \cite{Warren1995},
we impose an isothermal temperature field $T$. This approximation is
valid because the thermal diffusivity of the alloy is about four
orders of magnitude larger than the solutal diffusivity. In addition,
for the considered undercooling, the growth takes place under
diffusion limited conditions.

Surface energy and kinetic anisotropy are incorporated into the model
by formulating orientation dependent gradient energy $a(\phi, \nabla
\phi)$ and kinetic factor $\omega(\phi, \nabla \phi)$
\begin{eqnarray}
  \label{eq:a}
  a(\phi, \nabla \phi) = \gamma |\nabla \phi|^2 &=& \gamma_0 \left( 1
  + \epsilon_c \cos(4 \theta) \right) |\nabla \phi|^2\\
  \label{eq:omega}
  \omega(\phi, \nabla \phi) &=& \omega_0 \left( 1 - \epsilon_k \cos(4
  \theta) \right),
\end{eqnarray}
where $\theta$ is the angle between the vector $\nabla\phi$ and the
$x$ axis.

Assuming the mass fluxes to be linear functions of the thermodynamic
driving forces $\mu_i = \frac{\partial f}{\partial c_i}$, the mass
balance equations for the three components are
\begin{equation}
  \label{eq:mass-balance}
  \frac{\partial c_i}{\partial t}
  = - \nabla \left(
  \sum_{j=1}^3
  L_{ij}(c,\phi) \nabla \frac{-\mu_j}{T}
  \right),
\end{equation}
with mobility coefficients given by
\begin{equation}
  \label{eq:mobility-coefficients}
  L_{ij}(c,\phi) = \frac{v_m}{R}
  D_i c_i
  \left(
  \delta_{ij} - \frac{D_j c_j}{\sum_{k=1}^3 D_k c_k}
  \right).
\end{equation}
The form of Eq.~(\ref{eq:mobility-coefficients}) allows different
values of the bare trace diffusion coefficients $D_i(\phi)$ for the
different components $i$ and satisfies the constraint $c_1+c_2+c_3=1$.

Numerical aspects of phase-field modeling have been discussed in
\cite{Numeric} including the conditions for spatial grid resolution in
comparison to the interface thickness.  The evolution
Eqs.~(\ref{eq:phase-field}) and (\ref{eq:mass-balance}) are solved
using a finite element method with a semi-implicit time discretization
on a nonuniform adaptive mesh having the highest order of spatial
resolution in the vicinity of the solid-liquid interface where the
gradients of the phase field and of the concentrations reach maximal
values. The adaptive grid refinement criterion has been defined
ensuring a minimum of 7--10 grid points in regions of the diffuse phase
boundary. To verify the origin of interface instability, simulations
with different time and spatial resolution have been performed leading
to the same growth morphology. Hence, it has been shown that the
interfacial instabilities are of Mullins-Sekerka--type and not
numerically induced.

The solidus and liquidus lines of the binary Ni-Cu and of the Ni-Cr
phase diagram can be constructed from the free energy,
Eq.~(\ref{eq:free-energy-density}), using the melting temperatures
$T_{\mathrm{Ni}}=1728$~K, $T_{\mathrm{Cu}}=1358$~K,
$T_{\mathrm{Cr}}=1465$~K, the latent heats
$L_{\mathrm{Ni}}=2350$~J/cm$^3$, $L_{\mathrm{Cu}}=1728$~J/cm$^3$,
$L_{\mathrm{Cr}}=1493$~J/cm$^3$ and the molar volume
$v_m=7.42$~cm$^3$. The values of the melting temperature and of the
latent heat for Cr are adjustable parameters in order to recover the
actual binary phase diagram in the given region of concentrations. The
above values lead to a partition coefficient $k_e=0.843$, to a
liquidus slope $m_e=-3.27$~K/at.\% and to the undercooling $\Delta
T_0=24.4$~K for the binary Ni$_{60}$Cu$_{40}$ system. Similarly, we
obtain $k_e=0.905$, $m_e=-2.08$~K/at.\% and $\Delta T_0=8.7$~K for the
binary Ni$_{60}$Cr$_{40}$ system on the corresponding equilibrium
phase diagram.

To accentuate the effect of solidification interval, we assume that
surface properties (surface energy density $\sigma$ and kinetic
coefficient $\mu$) do not depend on alloy composition and have the
values $\sigma= \gamma_0 T =0.37$~J/m$^2$ and $\mu=3.3$~mm/(s$\cdot$K)
\cite{Warren1995}. The anisotropy of the interface properties plays an
important role in the selection of the operating state during
dendritic growth. In this study, we use the values obtained from
molecular-dynamics simulations for the anisotropies in pure Ni,
Eqs.~(\ref{eq:a}) and (\ref{eq:omega}), $\epsilon_c=0.023$ for surface
free energy density and $\epsilon_k=0.169$ for the kinetic coefficient
\cite{Hoyt1999}.  As reported in \cite{Hoyt1999}, the diffusion
coefficients of Ni and Cu in melt are $D_{\mathrm{Ni}}=3.82\times
10^{-9}$~m$^2$/s, $D_{\mathrm{Cu}}=3.32\times 10^{-9}$~m$^2$/s and we
assume $D_{\mathrm{Cr}}=1.5\times 10^{-9}$~m$^2$/s.  The diffusion
coefficient in the solid phase is set equal to $10^{-13}$~m$^2$/s for
all components.  The value of the small length scale parameter in the
entropy functional, Eqs.~(\ref{eq:entropy_functional}) and
(\ref{eq:phase-field}), is chosen to be $\varepsilon=0.1$~$\mu$m.

\begin{figure*}[t]
  \includegraphics[scale=0.9]{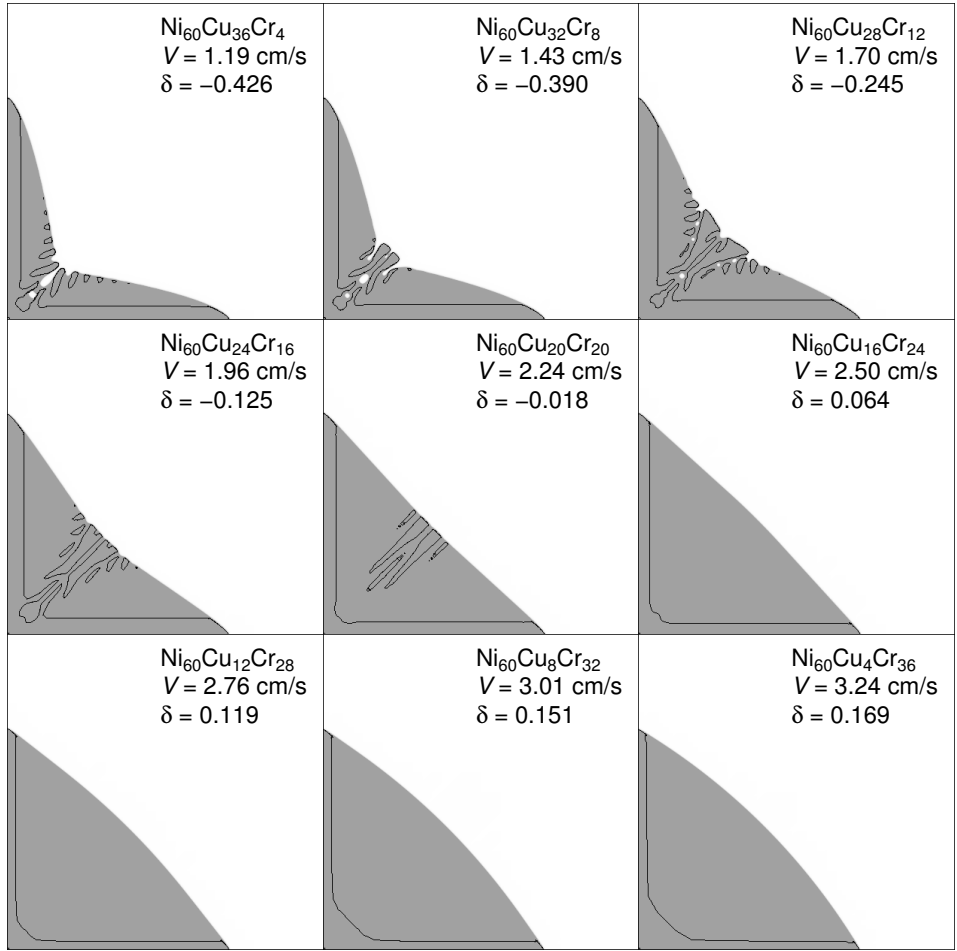}
  \caption{Transition from dendritic to globular
    morphology for different initial melt compositions at a constant
    initial undercooling of $20$~K. Shaded regions correspond to the
    solid phase, i.e. $\phi \geqslant 0.5$, and solid lines represent
    the isolines of average concentration of Ni in solid. The values
    for the tip velocities $V$ and for the area estimations $\delta$
    are stated in each image.}
  \label{fig:1}
\end{figure*}

%%%%%%%%%%%%%%%%%%%%%%%%%%%%%%%%%%%%%%%%%%%%%%%%%%%%%%%%%%%%%%%%%%%%%%
%%% Results %%%%%%%%%%%%%%%%%%%%%%%%%%%%%%%%%%%%%%%%%%%%%%%%%%%%%%%%%%
We conducted numerical computations for different alloy compositions
varying from Ni$_{60}$Cu$_{36}$Cr$_{4}$ to Ni$_{60}$Cu$_{4}$Cr$_{36}$
and for fixed undercooling conditions.  We kept the concentration of
Ni at 60\,at.\% and adjusted the initial undercooling at $20$~K
measured from the equilibrium liquidus line for a given composition.
The corresponding shapes of growing crystals are shown in
Fig.~\ref{fig:1}.  At small contents of Cr, the solid phase forms a
dendritic structure that has a pronounced preferable growth direction
determined by the anisotropy of surface energy and interface kinetics
(first three images of Fig.~\ref{fig:1}). An increase of Cr
concentration in the melt leads to an increase of dendritic tip
velocity $V$. This is accompanied by a thickening of the primary
dendritic trunk.  Further rise of Cr concentration (middle three
images of Fig.~\ref{fig:1}) causes a morphological transition from a
dendritic to a globular growth shape at a composition
Ni$_{60}$Cu$_{20}$Cr$_{20}$. The globular form of the crystals is
stable after this morphological transition for further change of alloy
composition to Ni$_{60}$Cu$_{4}$Cr$_{36}$.  Within the globular regime
of solidification, the anisotropy of the interface has a smaller
effect on the crystal shape.  The preferred growth directions are less
pronounced in comparison with the dendritic shape for amounts of Cr
less than 20\,at.\%.  To investigate the dependence of the spatial
redistribution of the alloy components in the solid phase on the melt
composition, we consider isolines corresponding to the average
concentration of nickel atoms in the solid phase. These isolines
separate Ni depleted and Ni enriched domains (see Fig.~\ref{fig:1},
solid lines). The character of the isolines for copper and chrome is
qualitatively the same only having small quantitative deviations. Two
independent parts can be found in the geometry of the isolines. The
first part is aligned in a direction parallel to the dendritic trunk
remaining in this position after the transition to the globular
morphology. The second part develops in diagonal direction and it
disappears at the morphological transition.

\begin{figure}
  \includegraphics[scale=1.0]{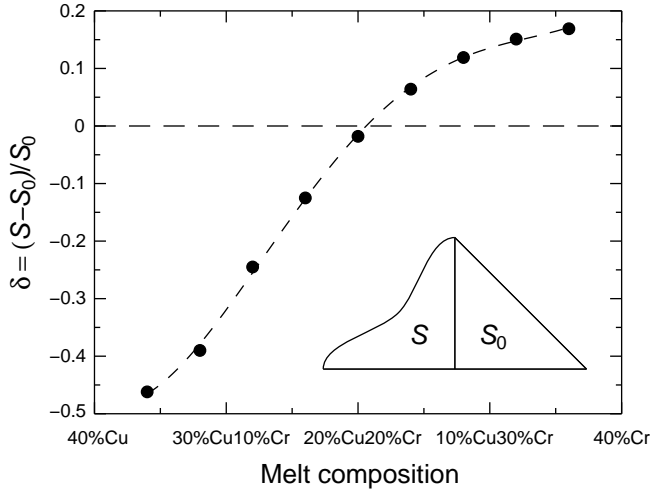}
  \caption{\label{fig:3} Simulated area $S$ covered by the crystal in
    relation to a normalized triangular area $S_0$ for different melt
    compositions. The $\delta =0$ line marks the transition from
    dendritic to globular growth morphologies.}
\end{figure}

The morphological transition can quantitatively be measured using the
deviation $\delta=(S-S_0)/S_0$ of area $S$ covered by the solid phase
from the triangular area $S_0$ between the dendritic tips and the
center of the initial seed, Fig.~\ref{fig:3}.  With the change of melt
composition, the deviation $\delta$ changes the sign from a negative
value for dendritic morphologies to a positive value for globular
shapes. The transitional value $\delta=0$ corresponds to the
composition Ni$_{60}$Cu$_{20}$Cr$_{20}$ and is hence in good agreement
with the transition point given by the vanishing diagonal part of the
isoline.  The transition takes place in a smooth manner.  On the one
hand, this follows from the linear increasing of the growth velocity
$V$ of the dendritic or globular tip with increasing Cr concentration
(see the values for $V$ in Fig.~\ref{fig:1}). On the other hand, the
smooth transition is confirmed by the smooth dependence of the
deviation $\delta$ on the melt composition, Fig.~\ref{fig:3}.  The
morphology of the growing structure is determined by the concurrence
of the stabilizing force due to the surface energy and of the
destabilizing force due to the concentration gradients on the
interface \cite{Mullins1964,Trivedi1986}. The morphological transition
can be explained considering the balance between the stabilizing and
destabilizing forces.  During the computed transition from a dendritic
to globular shape for varying composition of the ternary alloy
Ni$_{60}$Cu$_{40-x}$Cr$_{x}$ with $0<x<40$, the liquidus slope and the
deviation of the equilibrium partition coefficient from unity
decrease.  This leads to a reduction of the destabilizing forces on
the interface. The globular growth shapes are becoming stable against
perturbations.  Although the phase-field simulations of the ternary
system were conducted for fixed undercooling, we remark that a similar
morphological transition from dendritic to globular growth shapes has
been observed in numerical simulations of dendritic growth in binary
alloys \cite{Galenko94}, where the initial undercooling acts as a
control parameter for the transition. In this case a morphological
transition occurs through a coalescence of side branches with
increasing of the undercooling.

%%%%%%%%%%%%%%%%%%%%%%%%%%%%%%%%%%%%%%%%%%%%%%%%%%%%%%%%%%%%%%%%%%%%%%
%%% Conclusions %%%%%%%%%%%%%%%%%%%%%%%%%%%%%%%%%%%%%%%%%%%%%%%%%%%%%%
In summary, our numerical simulations show the occurrence of a
morphology transition from dendritic to globular growth structures in
the ternary Ni-Cu-Cr system for a varying alloy composition.  The
results demonstrate the general potential of the phase-field approach
in modeling multicomponent solidification in a thermodynamically
consistent way. Together with previous developments
\cite{Boettinger2002}, the work in this Letter shows that the diffuse
interface formulation provides a uniform description for modeling
complex microstructure evolution in pure, binary and multicomponent
alloy systems.

%%%%%%%%%%%%%%%%%%%%%%%%%%%%%%%%%%%%%%%%%%%%%%%%%%%%%%%%%%%%%%%%%%%%%%

% If you have acknowledgments, this puts in the proper section head.
\begin{acknowledgments}
  We thank P. K. Galenko from DLR Cologne for helpful discussions and
  gratefully acknowledge the financial support provided by the German
  Research Foundation (DFG)
  %within the priority research
  %program ``Phase Transformations in Multicomponent Melts'' 1120
  under Grant Nos. Ne 882/2-1 and Ne 882/2-2.
\end{acknowledgments}

%%%%%%%%%%%%%%%%%%%%%%%%%%%%%%%%%%%%%%%%%%%%%%%%%%%%%%%%%%%%%%%%%%%%%%
%% Bibliography %%%%%%%%%%%%%%%%%%%%%%%%%%%%%%%%%%%%%%%%%%%%%%%%%%%%%%
% Create the reference section using BibTeX:
%\bibliography{bibliography}

%%%%%%%%%%%%%%%%%%%%%%%%%%%%%%%%%%%%%%%%%%%%%%%%%%%%%%%%%%%%%%%%%%%%%%
%% End %%%%%%%%%%%%%%%%%%%%%%%%%%%%%%%%%%%%%%%%%%%%%%%%%%%%%%%%%%%%%%%
\end{document}